\documentclass{article}
\usepackage{amsmath}
\usepackage{enumitem}
\usepackage{multirow}
\usepackage{booktabs}
\usepackage{tabularx}
\usepackage{a4wide}
\usepackage{graphicx}
\usepackage{authblk}
\usepackage{cite}

\setlist[itemize]{topsep=4pt,itemsep=0mm}

\begin{document}

\title{Microsimulations of demographic changes in England and Wales under different EU referendum scenarios}
\author{Werpachowska, A.~M.\footnote{\texttt{aw@averisera.uk}, Averisera Ltd, London, United Kingdom}\ \ and Werpachowski, R.}
\maketitle

\begin{abstract}
We perform stochastic microsimulations of the dynamics of England and Wales population after the British referendum on EU membership, considering different possible outcomes. Employing available survey data, we model the demographics of the region over the next generation, as shaped by births, deaths and international migration. The migration patterns between England and Wales and the remaining EU countries are modified according to the possible scenarios of their future relations. We find that Brexit will accelerate the overall population ageing and the deepening imbalance between workers and retirees but reduce the population growth and the fraction of women of reproductive age. In the alternative scenarios of remaining in the EU these effects will be partially forestalled by the influx of immigrants from current and prospective EU countries and their children. In all considered scenarios the native British population declines. Our study demonstrates that microsimulations can be a useful tool for designing and evaluating the country's policies in the advent of fundamental transformations.
\end{abstract}

\section{Introduction}

The debate over Britain's possible exit from the European Union (Brexit) in the run-up to the referendum on 23 June 2016 has brought up several questions about the changing demographics of the country and their socioeconomic implications. The rich multiethnic, multicultural structure of the UK society has been historically shaped by several waves of economically motivated migration, the last of which began after the accession of Eastern European countries to the EU in 2004. Similarly, the referendum decision will influence its future composition through population shifts in migration patterns between Britain and the rest of the EU.

In this work we compare the impact of different policy scenarios on the England and Wales (E\&W) demographics, covering different referendum outcomes and subsequent relationships with the EU described in the next section. To this end, we perform microsimulations~\cite{Orcutt,Morand,Zagheni,Duleep,Andreassen} of a representative sample of E\&W population (stratified by age, sex and ethnicity) tracking individual histories (births, pregnancies, path-dependent migration and deaths) of its members from 1991 to 2041 and varying their migration patterns after Brexit. The initial state and stochastic processes driving its dynamics are calibrated against data taken from the UK Censuses, birth and death registrations, International Passenger Survey and Labour Force Survey conducted by the Office for National Statistics (ONS), as described in detail in Sec.\,\ref{sec:mod-quant}. We analyse the implications of the scenarios for the future age, sex and ethnic structure of E\&W population, and estimate its basic socioeconomic indicators in Sec.\,\ref{sec:results}.

\section{Microsimulation scenarios}
\label{sec:scenarios} 

The five simulated scenarios depend predominantly on whether the UK remains in or leaves the EU and affect the migration patterns of native British and EU immigrants as follows:

\noindent\textbf{Remain scenarios}
\begin{itemize}
	\item \textbf{Status quo}\ extrapolates current population and migration trends.
	\item \textbf{2nd enlargement}\ describes the UK remaining in the EU and subsequent additional inflow of immigration after the accession of new countries.
\end{itemize}

\noindent\textbf{Brexit scenarios}
\begin{itemize}
	\item \textbf{Soft Brexit}\ assumes an amicable parting of the ways between Britain and the EU. Their relations do not change significantly and the migration flows experience only slight changes.
	\item \textbf{Exodus}\ envisions a significant ebb of EU immigrants from the UK, with an accompanying slight drop in British emigration from the country and a minor increase of their repatriation from the EU.\footnote{A comparison of this scenario with the Remain scenarios can test the common perception among the Leave campaigners that if the UK remains in the EU its population will increase by 2--3 million by 2030.}
	\item \textbf{Hard Brexit}\ leads to drastically limited migration between the UK and the EU after Brexit as many migrants lose the right of residence or decide to return to their country of origin.
\end{itemize}

Each scenario involves the following migration flows to and from the country:

1) \textbf{Inflow of EU immigrants}\quad 
The outcome of the Brexit referendum and further EU enlargement affect the migration flows of non-British EU citizens. We assume that they will return to levels from before 2004 after Brexit takes place in 2019 (i.e.~2 years after formally announcing the intention to withdraw from the Union, in accordance with the Treaty on European Union). If it is otherwise, they will maintain their current trends to the end of simulation; additionally, if new countries (including Turkey and Western Balkans) join the EU by the oft-cited year of 2020, a new wave of immigration will occur, raising the current migration flows between E\&W and the remaining EU countries by a factor $f_\text{Enl}$ for 10 following years.

2) \textbf{Outflow of EU immigrants}\quad
After Brexit, a part of the EU immigrants living in E\&W will return to their country of origin. We expect that this `exodus' will start 2 years after the referendum and last another 2 years. Depending on the scenario different fraction $f_\text{Ex}$ of this group will leave during this period. Additionally, those who will have arrived closer to the Brexit date will leave first (last-in-first-out instead of random selection). The exodus will not involve the immigrants arriving after Brexit.

3) \textbf{Outflow of native British to the EU}\quad 
About 30\% of British emigration is destined for other EU countries~\cite{IPS}, either for work or retirement. We assume that this outflow will be reduced by Brexit by a factor $f_\text{Em}$.

4) \textbf{Inflow of native British from the EU}\quad
Brexit may start a wave of returns of British emigrants living in other EU countries. We model it by increasing their repatriation to E\&W by a factor $f_\text{Ret}$. We assume that the process will begin 2 years after the referendum and last 2 years.

The values of the simulation parameters for the above scenarios (Table~\ref{tab:scen-params}) are surmised from the ONS data, in particular the International Passenger Survey. They are discussed in detail in Sec.\,\ref{sec:mod-migration}.

\begin{table}[h]%
\centering\begin{tabular*}{24em}{p{9em} @{\extracolsep{\fill} } c c c c}
\toprule
 & $f_\text{Enl}$ & $f_\text{Ex}$ & $f_\text{Em}$ & $f_\text{Ret}$ \\[0.24cm]
Status quo & 100\% & 0\% & 100\% & 0\% \\
2nd enlargement & 200\% & 0\% & 100\% & 0\% \\[0.14cm]
Soft Brexit & 0\% & 10\% & 80\%  & 10\%\\
Exodus & 0\% & 70\% & 80\% & 10\% \\
Hard Brexit & 0\% & 70\% & 30\% & 80\% \\
\bottomrule
\end{tabular*}
\caption{Microsimulation parameters for post-referendum scenarios.}
\label{tab:scen-params}
\end{table}

\section{Methods and data}
\label{sec:mod-quant}

\subsection{Microsimulation engine}

Our microsimulation engine employed in this study generates the stochastic evolution of a complex system through path-dependent dynamics of individual agents controlled by initial conditions, different values of attributes and exogenous factors. It supports the dynamic addition and removal of agents (such as births and deaths of persons), their transitions (e.g.\,migration between the EU and E\&W), time-dependent interactions between agents (e.g.\,links between mothers and children), memory (e.g.\,last-in-first-out migration or course of pregnancy), as well as other features not used in this project, such as multi-level simulations, correlated and dynamically changing control variables, or coupling to a reservoir with a possible feedback mechanism. The flexible engine design facilitates building microsimulation models for forecasting, scenario analysis and intervention testing. The numerical code is written in modern C$^{_{\text{\tiny++}}}$ and runs on Windows and Linux. It provides an object-oriented application programming interface (API) which can be used to independently implement additional microsimulation features.

\subsection{Microsimulation model for forecasting the E\&W demographics under different EU membership scenarios}

We have built the stochastic dynamic microsimulation model for forecasting changes of the E\&W demographics under different scenarios of future relations with the EU. It combines the headship rate method (enforcing a given distribution, conditional or not, of a particular variable in the population at given time; used to model birth multiplicity and sex of newborn children)~\cite{Andreassen} and transition matrix method (applying transition probabilities to change the state of an individual member; used to model e.g.\,mortality)~\cite{Leslie,Andreassen} to model the evolution of the population. Since our model describes the dynamics of many cohorts at the same time, it can be classified as a cross-sectional dynamic microsimulation~\cite{Andreassen}. Compared with other popular dynamic demographic models~\cite{WertheimerII,Hammel1976,GallerWagner,Wolfson,NelissenVossen,Nelissen,Andreassen}, ours is quite simple: focusing on natural growth and migration, it does not describe any economic variables (income, wealth or labour force participation) or educational attainment, while the household formation is only partially addressed as mother-child links. The treatment of migration is similar to the one in the MOSART model~\cite{Andreassen} with additional stratification by ethnic group. The structure of our model and the data it uses are described in the following parts of this section. 

We model the E\&W population from 1 July 1991 to 1 July 2041 with a quarterly time step. The initial sample contains 5 million persons, which is approximately 10\% of the population as measured by the 1991 UK Census. Each of its members is characterised by fixed attributes, i.e.~date of birth (the daily precision minimises the time discretisation error), sex (male or female) and ethnicity (one of 18 ethnic groups according to the ONS classification used in the 2011 UK Census). Their values are drawn randomly from the three-dimensional distribution of sex, age (tabulated in five-year groups: 0-4, 5-9, \dots, 85 and over) and ethnicity reported by the Census. The early start date serves to generate realistic life histories of individual persons, in particular building up immigrant and emigrant stock in preparation for Brexit modelling (the starting sample, while reflecting accurately the sex, age and ethnic structure, does not contain any immigrants, emigrants or mothers).

Life events, such as births, pregnancies, migration and deaths, are described by stochastic processes. Their parameters are calibrated to the following ONS data sources: birth rates, birth multiplicities and child sex ratios to E\&W births registrations~\cite{ONSChildbearing,ONSBirthSummaryTables,ONSBirthchar}, ethnicity adjustment factors for birth rates to Labour Force Survey~\cite{LFS}, and mortality rates to E\&W deaths registrations and population estimates~\cite{ONSDeathsByAge,ONSPopEst}. Migration rates are calculated based on UK Censuses (1991, 2001 and 2011)~\cite{ONS} and compared against International Passenger Survey~\cite{IPS} as sanity check. For the purpose of this study, we identify the population of native British with White British and Irish groups of the Census ethnic classification and the immigrants from current and future EU accession countries with Other White group (thus neglecting the impact of Brexit on the immigration from outside of the EU, the assumption we take into account when choosing the values of factors modifying the migration parameters in different scenarios in Table\,\ref{tab:scen-params}). The analysed microsimulation scenarios alter the migration patterns between E\&W and the EU as described in Sec.\,\ref{sec:scenarios}.

The full history of each person (including links between mothers and children) is by default preserved in memory throughout the simulation run, enabling us to model path-dependent demographic mechanisms with arbitrarily long memory or retrieve any additional information. We save to disk the numbers of all life events and the size of each sex, age in five-year intervals and ethnicity group. A single run for a sample simulated in this model takes approximately 10 hours on a standard PC and requires 4\,GB of RAM. The following parts of this section provide more modelling details.

\subsection{Births}
\label{sec:births}

We model births as resulting from pregnancies which are triggered by a non-homogeneous Poisson process with memory, with intensity (conception rate) dependent on the woman's age, year of birth and time since last childbirth (the probability of becoming pregnant again is set to zero for three months after giving birth). For simplicity, every pregnancy is assumed to be successful. The conception rate is calibrated to historical data on birth rates by woman's age (in years, from 15 to 45) and year of birth of woman~\cite{ONSChildbearing} and multiplicity (one or two) by five-year age group and year~\cite{ONSBirthchar}. If birth rate for a given year of birth and age is yet unknown (e.g.~women born after 1999), we use the latest available birth rate for women of the same age. Next the conception rate for each major E\&W ethnic group (White British, Other White, Indian, Pakistani, Bangladeshi, Chinese, African and Caribbean) is rescaled by the time-dependent ratio of total fertility rate (TFR) for this ethnic group to the average TFR of the whole population (the TFR data are extrapolated flat forward)~\cite{DynamicsDiversity, Coleman}. We assume that TFR for the remaining ethnic groups is equal to the average (since they comprise less than 4\% of the total population, the error introduced in this way is small).

The simulation preserves the information about the mother-child relationship. New children are assigned mother's ethnic group, while their sex is randomly drawn from a time-dependent two-point distribution based on historical data on the number of male and female births~\cite{ONSBirthSummaryTables}. 


The employed pregnancy model, extensible to general epidemiological applications, enables us to generate more realistic childbearing histories and, in consequence, reduce the error of our results. Assuming an effective minimum one year interval (9 months of pregnancy and 3 months of postpartum infertility) between births, we reduce the variance of their number per woman. Modelling births as a simple Poisson process with arbitrarily small intervals between births would lead to the same mean, but a significantly higher variance, and hence larger Monte Carlo errors in the simulation.

\subsection{Mortality}

Mortality is modelled as a non-homogeneous Poisson process with mortality rate dependent on sex, age and year of birth. Thus, differences in mortality between ethnic groups result from their different sex and age structure. We calibrate the mortality curves, i.e.~mortality rates as a function of age for given year of birth and sex of person, to annual data on mortality rates by sex and five-year age group~\cite{ONSDeathsByAge,ONSPopEst}. If all or part of data points are not available for any cohort, we copy them from the closest cohort for which the data are known (e.g.~if the last known data on mortality rate of 80-year-old men is for a cohort born in 1930, we use this mortality rate also for men born in 1931). Differently to some other authors~\cite{Andreev} and consistent with recent observations~\cite{ONSAvoidMort}, we do not assume that the decline in age-specific mortality rates observed so far will continue indefinitely in the future. Mortality rate for a given cohort is a piecewise constant function of age with five-year intervals until the age of 90, beyond which it is extrapolated flat.

\subsection{Migration}
\label{sec:mod-migration}

Given the lack of migration data for different ethnic groups in the E\&W population,\footnote{International Passenger Survey measures immigration/emigration, source/destination countries and
the nationality of migrants, but does not record their ethnicity. Therefore, we cannot combine it with UK Censuses---our main and the most complete source of data on the demographics of E\&W population, which measure ethnicity, but not nationality. For simplicity, we calibrate our model to net migration calculated based on the Census data, adapting the analytical method used in Ref.\cite{DynamicsDiversity}. It is worth noting that the net migration figures obtained from our microsimulation agree with the results of the cited work, e.g., for White British group we obtain the net migration in years 2001--2011 of -1.47\% of the initial (2001) population compared to -1.51\% in the cited work and for Indian group we obtain 22.91\% compared to 21.5\%.} we model it based on the decennial census data from 1991, 2001 and 2011~\cite{ONS} in the following way:
\begin{enumerate}
\item\label{itm:simnomigr} For each decade $[y, y+10]$ between censuses, where $y=$ 1991 or 2001, we carry out the simulation without any migration, and record the population composition by sex, age and ethnicity on the 1 July of years $y$ and $y+10$. Next, we rescale each recorded population by the ratio of the historic to simulated E\&W population size from 1 July $y$.

\item For each five-year birth cohort $i$ with given sex and ethnicity in census year $y$ we collect the following quantities:
\begin{itemize}
	\item[] $n_{yi}$ -- the size of cohort $i$ obtained from the simulation in step 1 in year $y$,
	\item[] $n'_{yi}$ -- the size of the corresponding group censused in year $y$,
	\item[] $n_{y+10,i}$ -- the size of cohort $i$ after 10 years,
	\item[] $n'_{y+10,i}$ -- the size of the corresponding group censused in year $y+10$.
\end{itemize}
\item We estimate the net migration based on the mid-year census data in the period between 1 July $y$ year and 1 July $y+10$ year in the above cohort using the formula $(n'_{y+10,i} - n'_{yi}) - (n_{y+10,i} - n_{yi})$. The positive sign indicates immigration and negative emigration.
\end{enumerate}

The net migration rate $k_i$ in each cohort can be modelled either as relative, $dn_i/dt = k_i n_i$, or absolute, $dn_i/dt = k_i$. We assume that the first is better suited for describing the emigration of dominant (native) ethnic groups from an economically and politically stable country, where no substantial push factors affect the probability of moving abroad. In contrast, the second describes more realistically the migration of ethnic minorities, which is mainly driven by sentiments or various factors in the country of origin (e.g.~the inflow of Polish workers after 2004). Accordingly, we use the relative rate to model the emigration of White British and Irish from E\&W and the absolute rate to model the emigration and immigration of the remaining, smaller ethnic groups (such as Other White), as well as the repatriation of White British and Irish. From the above formulas, we obtain the relative net migration rate $k_i = {\log (1 + \Delta m_i / n_i)}/{\Delta t}$ and the absolute rate $k_i = {\Delta m_i}/{\Delta t}$ for each cohort $i$, where $n_i$ is the size of the cohort at the beginning of $\Delta t$ period and $\Delta m_i$ is a net migration level over time $\Delta t$.

Simulating the net migration may obscure the true scale, structure and dynamics of migration flows in the population (e.g.~a large number of emigrants may offset a large number of immigrants), as noted in e.g.\cite{ODonoghue}. We partially ameliorate these problems by using a fine stratification of cohorts (by age in five-year intervals, sex and ethnicity) when calibrating migration rates.

In the simulation, we apply the obtained $k_i$ values to each cohort over time $\Delta t$, to calculate $\Delta m_i = n_i (e^{k_i \Delta t} - 1)$ or $\Delta m_i = \max(k_i \Delta t, -n_i)$ in this group for relative and absolute rates, respectively. Beyond the last census date, extant and new persons are split in groups by five-year age intervals, sex and ethnicity, and assigned the net migration rates of corresponding groups from year 2006 (halfway between the 2001 and 2011 Censuses) in the ``Status quo'' scenario. We modify these rates in other scenarios described in Sec.\,\ref{sec:scenarios}.

Emigrants and immigrants are selected randomly from each five-year birth, sex and ethnicity cohort. White British and Irish emigrants are moved to an auxiliary emigrant population, which is simulated in parallel to the E\&W population using the same assumptions about births and mortality, but without further migration; it serves to simulate a more realistic age and sex structure of the native British emigrants returning to E\&W in the simulated post-Brexit scenarios. Immigrants are cloned to create new persons who are added to the E\&W population. All children below the age of 10 migrate together with their mothers.

We modify the above migration dynamics by factors from Table\,\ref{tab:scen-params} to account for different post-referendum scenarios. The ``Status quo'' case extrapolates obtained trends without any changes. In the ``2nd enlargement'' the net migration levels of Other White are multiplied by the $f_\text{Enl}$ parameter between 2020 (the next considered EU enlargement by countries with Other White majority) and 2030, reflecting the increase in the migration flows in this group, and then come back to the state from 2016. In all Brexit scenarios, starting from the Brexit date (set to 1 July 2019, which is the closest date in the microsimulation schedule to the official one), the emigration of White British and Irish drops by the factor $f_\text{Em}$, while the fraction $f_\text{Ret}$ of the auxiliary population of native British emigrants returns to E\&W over the following two years. (Since the available data force us to work with net migration rather than separate immigration and emigration figures, it is important to note that changes of emigration rates generally correspond to larger relative changes of net migration, possibly even changing its sign. For this reason, the $f_\text{Em}$ values must be very high to account for realistic shifts in emigration. Additionally, $f_\text{Ret}$ must be increased to compensate for the fact that the stock of native British emigrants to the EU accumulated in the microsimulation course does not include people who emigrated before 1991.) At the same time, the net migration of Other White returns to the levels from 2000 (before the 2004 EU enlargement) and the exodus of this group begins, during which the fraction $f_\text{Ex}$ of this group leaves E\&W over the following two years. The exodus is modelled using the last-in-first-out algorithm. It assumes that the newest EU immigrants, who have spent less time in E\&W, have not yet formed strong bond with their new country and kept more ties with their country of origin, which makes them more willing and able of all EU immigrants to return. Furthermore, any right to remain granted by the UK government to EU immigrants after Brexit is likely to depend on how long they have been residing in the country.

\subsection{Error and stability analysis}
\label{sec:sampl-err}

\paragraph{Sampling error}

A single person in the simulated population represents approx.~10 real persons in E\&W. This sample size reduction causes the results to be less reliable for small groups (e.g.~70-75-year-old Chinese males). To estimate the sampling error for the fractional values (size of a group relative to the total population), at each future time $t$ we assume that the observed fraction of people belonging to a given group is drawn from a multinomial distribution. The parameters of this distribution are estimated within a Bayesian framework, using a flat Dirichlet prior and assuming that each observation (a person) is independent (this is not strictly true, because the last-in-first-out migration algorithm used in migration modelling and the migration of children together with mothers introduces correlations between population members, but it simplifies the estimation of the sampling error). The relative sampling error for ethnic, age and sex groups is negligibly small thanks to the large sample size (5 million). Estimation of the relative sampling error for the total population size would require performing bootstrapping (repeating the simulation run 200--300 times), which is not economical. However, we expect this error to be negligible owing to the large size of the simulated sample.

\paragraph{Sensitivity to input parameters}

\begin{figure}[h!]
\centering
\includegraphics[width=0.8\linewidth]{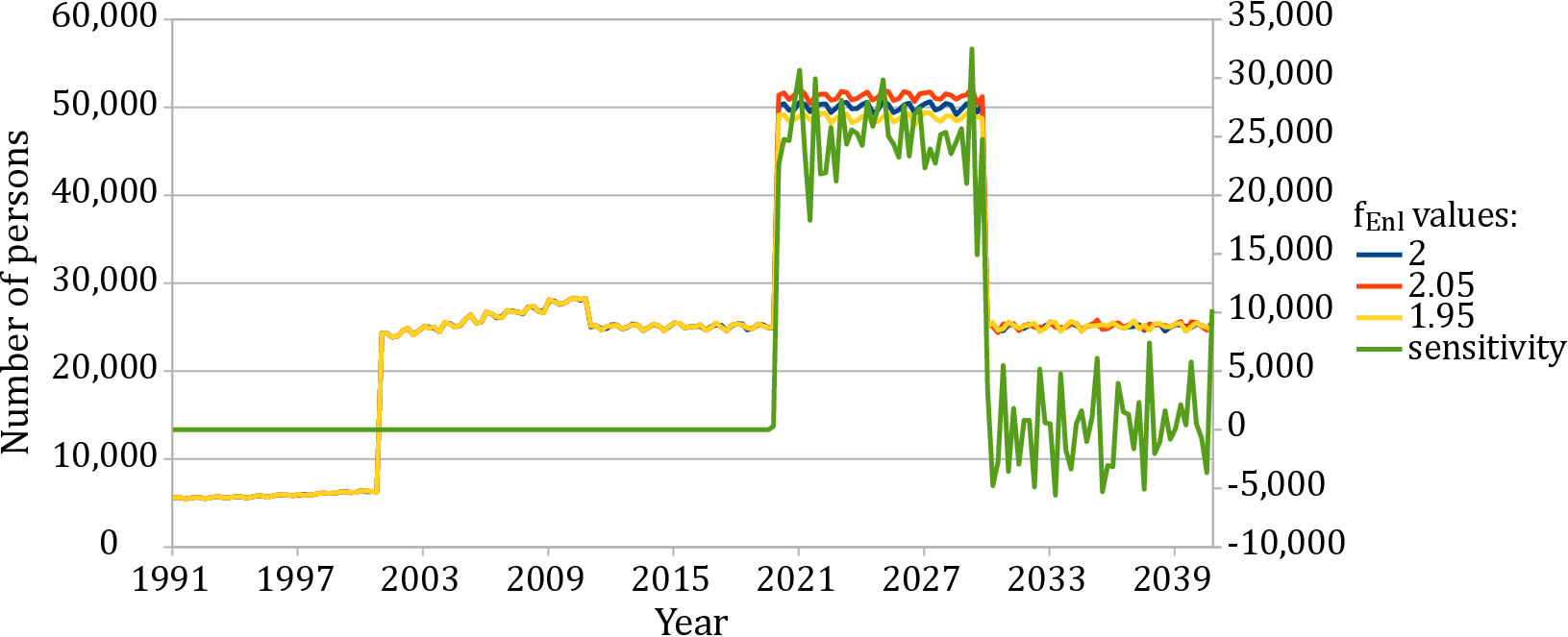}
\caption{Sensitivity (two-sided numerical derivative for $\pm$5\% perturbation, right axis) of quarterly immigration flow (left axis) for Other White ethnic group to $f_\textrm{Enl}$ parameter in the ``2nd enlargement'' microsimulation scenario. The wiggles result from the Monte Carlo sampling error.}
\label{fig:sensitivity}
\end{figure}

We carry out sensitivity analysis with respect to the scenario parameters $f_\text{Ex}$, $f_\text{Em}$, $f_\text{Ret}$ and $f_\text{Enl}$ by perturbing them up and down by 5\%. We expect the impact of these parameters on the final results to be approximately linear. Since $f_\text{Em}$ and $f_\text{Ret}$ affect the same segment of the population (White British and Irish), they need to be perturbed both separately and together to test the linearity assumption. An example of sensitivities to the microsimulation parameter $f_\text{Enl}$ is presented in Fig.\,\ref{fig:sensitivity}. The chart shows that the immigration flows in years 2020--30 can be accurately controlled by this parameter as the sensitivity to it is precisely zero before this period and then starts to fluctuate around zero as a result of the Monte Carlo error.

We do not combine the sampling errors, calculated as described in the previous section, with the error resulting from the sensitivity to (unknown precisely) scenario parameters, because our goal is to illustrate a range of possible outcomes, not to predict the most likely one.

\section{Results}
\label{sec:results}

\subsection{Total size of the E\&W population}

\begin{figure}[h!]
\centering
\includegraphics[scale=0.51,trim=10px 30px 30px 10px]{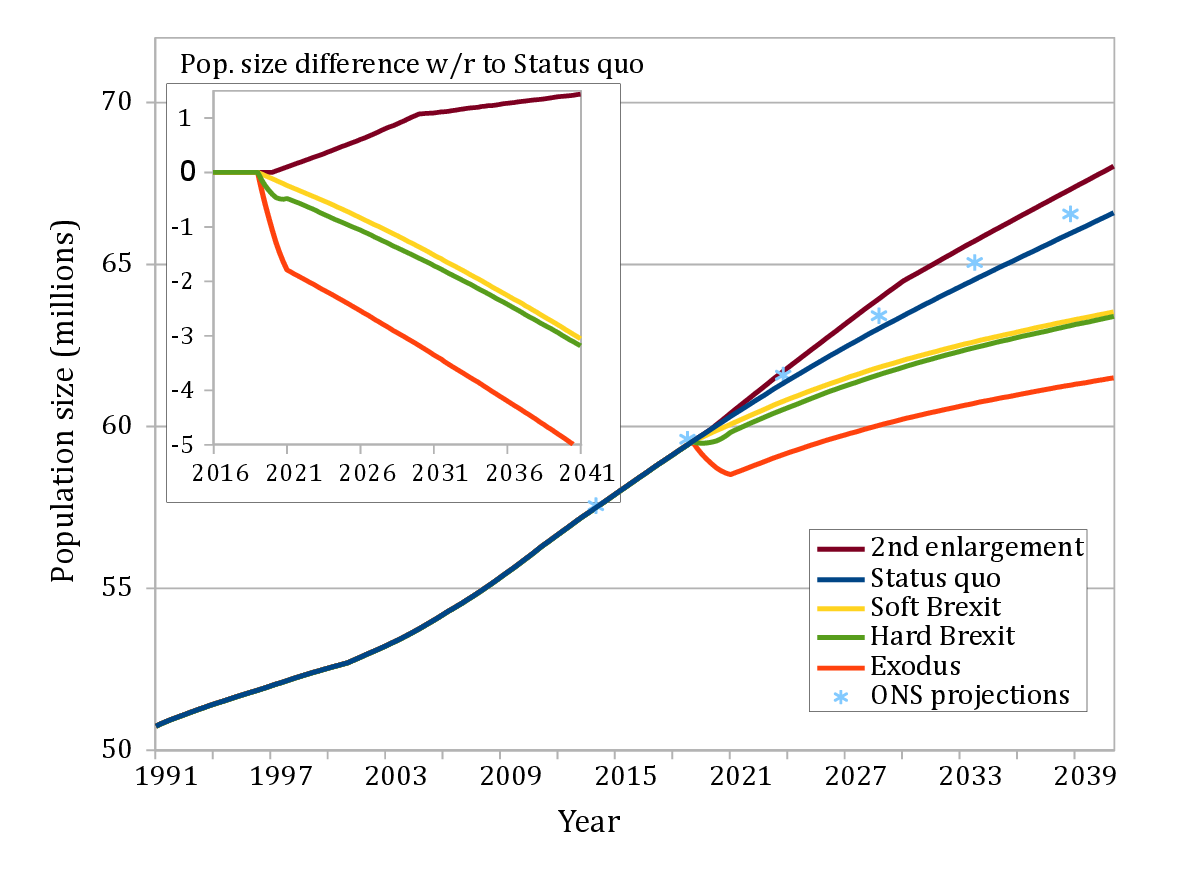}
\caption{Total size of the E\&W population under different EU membership scenarios and the ONS projections~\cite{ONSprojections}. The dip after 2019 in the ``Exodus'' scenario is caused by the large-scale outflow of the EU immigrants. Inset: differences in the population size between the ``Status quo'' and other scenarios.}
\label{fig:tot_pop}
\end{figure}

Differences in population size between the considered scenarios become apparent within a few years, as shown in~Fig.\,\ref{fig:tot_pop}, confirming the significant influence of Britain's EU membership on its demographic structure. The E\&W population growth is fuelled by extending lifespan and immigration along with the higher fertility rate of immigrant women. Brexit will abate, but not halt it (see insert). The ONS estimates obtained by a static extrapolation called ``cohort component method''~\cite{ONSprojections}, included for comparison, are slightly higher than our ``Status quo'' prediction.

\subsection{Median age}
\label{sec:median-age}

\begin{figure}[h!]
\centering
\includegraphics[width=1\linewidth]{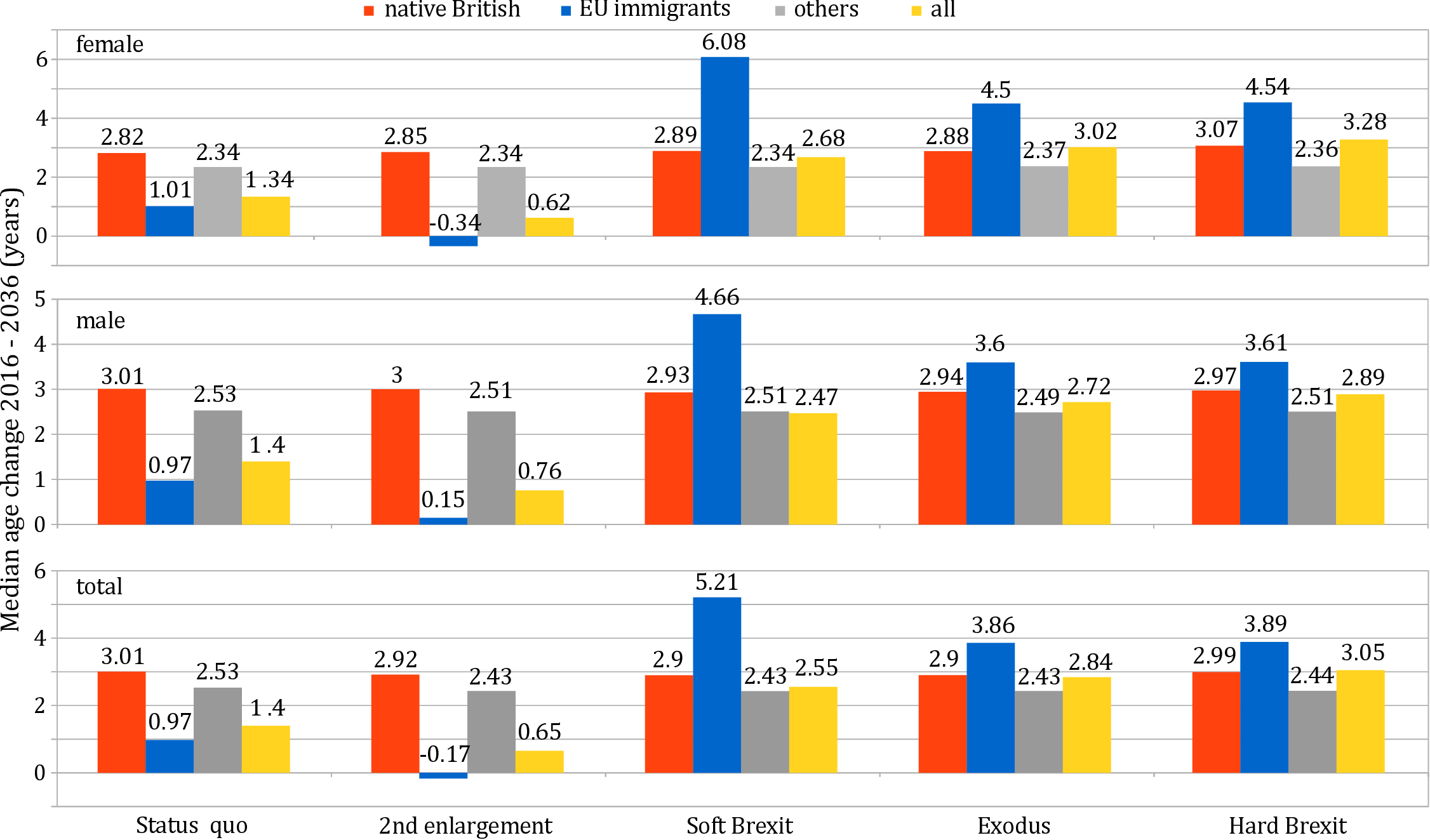}
\caption{Change in the median age of female, male and total E\&W population between 2016 and 2036 under different scenarios for native British, EU immigrants and the remaining ethnic groups (``others'').}
\label{fig:median_age}
\end{figure}

The future median age of E\&W population rises in all simulated scenarios, as presented in Fig.\,\ref{fig:median_age}. The trend concerns almost all sex and ethnic groups, except the EU immigrants in the ``2nd enlargement'' case, where it is inverted by the wave of young immigrants from new accession countries. In all Brexit scenarios the median age of the EU immigrant population is expected to increase sharply, as those who arrived last to E\&W and thus are more likely to return to their country of origin are on average younger than the earlier immigration and, at the same time, the inflow of new EU immigrants is reduced. Consequently, this segment of the E\&W population, which was unusually young in 2016, will see its median age rise much faster than other ethnic groups, which were already significantly older.

Since women constitute 53-55\% of the EU immigrant population (in particular, 52.5-56.5\% of the 0-29 age group), the post-Brexit immigrant exodus will affect the female median age more than the male one. This effect will be enhanced by the sex gap in life expectancy. Alongside the declining fertility rates, the departure of young female EU immigrants at the peak of their reproductive age will further accelerate the E\&W population ageing through a declining birth rate.

\subsection{Women of reproductive age and male to female ratio}
\label{sec:women-reproductive}

\begin{figure}[h!]
\centering
\includegraphics[width=1\linewidth]{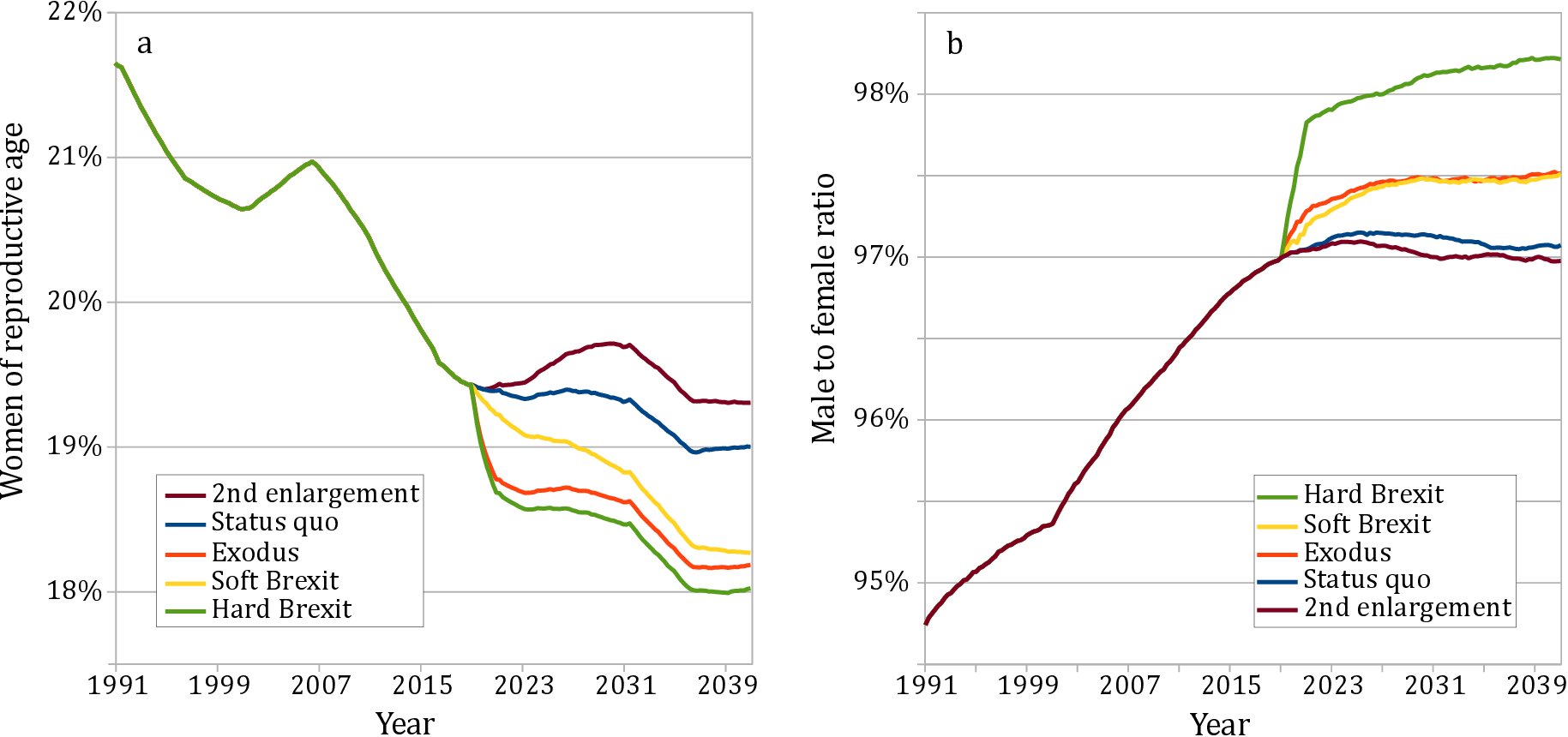}
\caption{a) Percentage of women of reproductive age in the total E\&W population and b) male to female ratio under the considered EU membership scenarios.}
\label{fig:women_reproductive}
\end{figure}

The percentage of women of reproductive age (15-45) in the E\&W population is expected to fall in all considered scenarios (Fig.\,\ref{fig:women_reproductive}a), which will negatively impact the natural growth. As pointed out in the previous section, Brexit reduces their number sharply because of the specific age and sex structure of the EU immigrant group, namely younger and with higher share of women than the E\&W average. Consequently, the male to female ratio in the total population will increase, as presented in Fig.\,\ref{fig:women_reproductive}b. The overall ageing of E\&W society imposes the same trends, although milder, on the ``Status quo'' scenario. They can be offset by higher rates of immigration after the future EU enlargement.

All scenarios reveal two periods in which the number of women of reproductive age and their share in the E\&W population increase: the opening decade of this century and years 2028--2036. The first is caused by the 2004 EU enlargement and the ensuing flow of immigrants from Eastern Europe, while the latter by their children entering the reproductive age. The resultant past and forthcoming small ``baby booms'' partially ameliorate the population ageing, without significantly encumbering the dependency ratio lowered by their parents (see Secs.\,\ref{sec:median-age} and~\ref{sec:dependency-ratios}).

\subsection{Dependency ratios}
\label{sec:dependency-ratios}

\begin{figure}[h!]
\centering
\includegraphics[width=1\linewidth]{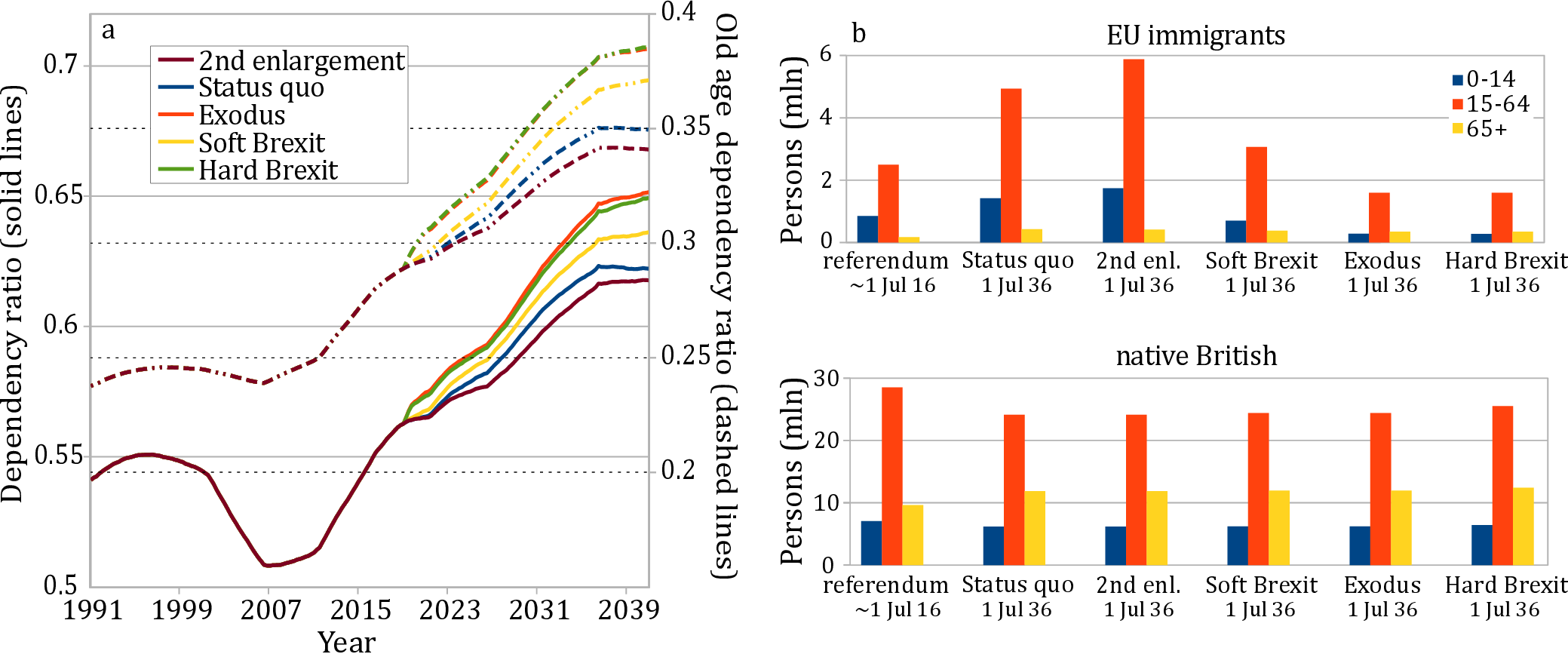}
\caption{Working-age group relative to children and elderly in the E\&W population: a) Dependency ratio (solid line, left axis) and old age dependency ratio (dashed line, right axis), as well as b) size of the age groups in the EU immigrant and native British populations days after the EU referendum and, following different EU membership scenarios, in 2036.}
\label{fig:workers_retirees}
\end{figure}

Figure~\ref{fig:workers_retirees}a displays the relationship between dependents (children and elderly) and the working-age group (ages 15-64). Their proportion, i.e.~the dependency ratio (solid line, left axis), exhibits a large dip from 2001 to 2011, caused by the simultaneous decrease in the fertility rate and the influx of EU workers. The latter also explains the flattening of the old age dependency ratio (of retirement-age to working-age persons; dashed line, right axis) in this period. After the influx subsides, and the female EU immigrants start having children, both curves pick up their previous upward trends. From 2019 onwards, the scenarios fan out, depicting differences between the migration patterns: Remain scenarios yield lower ratios than Brexit ones, in particular the dependency ratio ranges from a low of 0.62 under the ``2nd enlargement'' scenario to a high of 0.65 under the ``Exodus'' scenario in 2041, whereas the old-age dependency ratio from 0.34 to 0.38. All curves flatten after 2036, when ``baby boomers'' (see Sec.\,\ref{sec:women-reproductive}) enter the workforce.


The divergence between the analysed scenarios is mainly driven by working-age EU immigrants (see Fig.\,\ref{fig:workers_retirees}b). From 2016 to 2036, their number doubles if E\&W remains in the EU, maintains a similar level after ``Soft Brexit'', and almost halves in the radical Brexit scenarios. The number of children of EU immigrants grows proportionately to the working-age group in Remain scenarios, but it shrinks faster than this group after Brexit as a result of the reduced inflow of young EU immigrant women; all scenarios predict a similar size of the elderly group through 2036.

Comparing the proportions of age groups in Fig.\,\ref{fig:workers_retirees}b shows that the EU immigrant population is characterised by lower dependency ratios than the native British one. The continued ageing of the latter (see Sec.\,\ref{sec:median-age}) additionally contributes to the overall growth of dependency ratios in the total E\&W population.

\subsection{Population pyramid in 2036}

\begin{figure}[h!]
\centering
\includegraphics[width=1\linewidth]{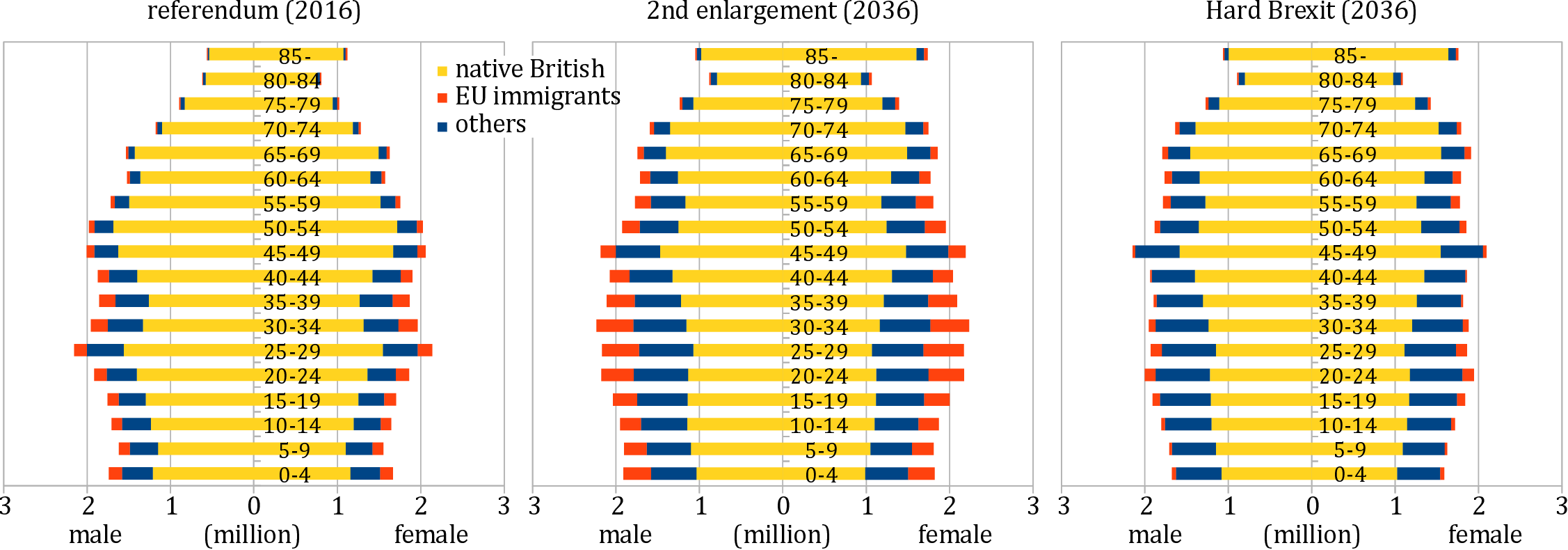}
\caption{Population pyramids for E\&W in 2016 and 2036 for two opposite EU membership scenarios (``others'' denotes the remaining population, except native British and EU immigrants).}
\label{fig:population_pyramids}
\end{figure}

Population pyramids for all considered EU membership scenarios have the shape characteristic of an ageing society, as shown in Fig.\,\ref{fig:population_pyramids} for the E\&W population in 2016 and 2036. In Remain scenarios they have broader lower halves than the Brexit ones owing to the influx of young immigrants. However, this broadening does not extend down to the lowest age groups, highlighting the difference between population change due to natural growth and immigration.

\subsection{Mechanisms of population change 2016--2036}

\begin{figure}[h!]
\centering
\includegraphics[width=1\linewidth]{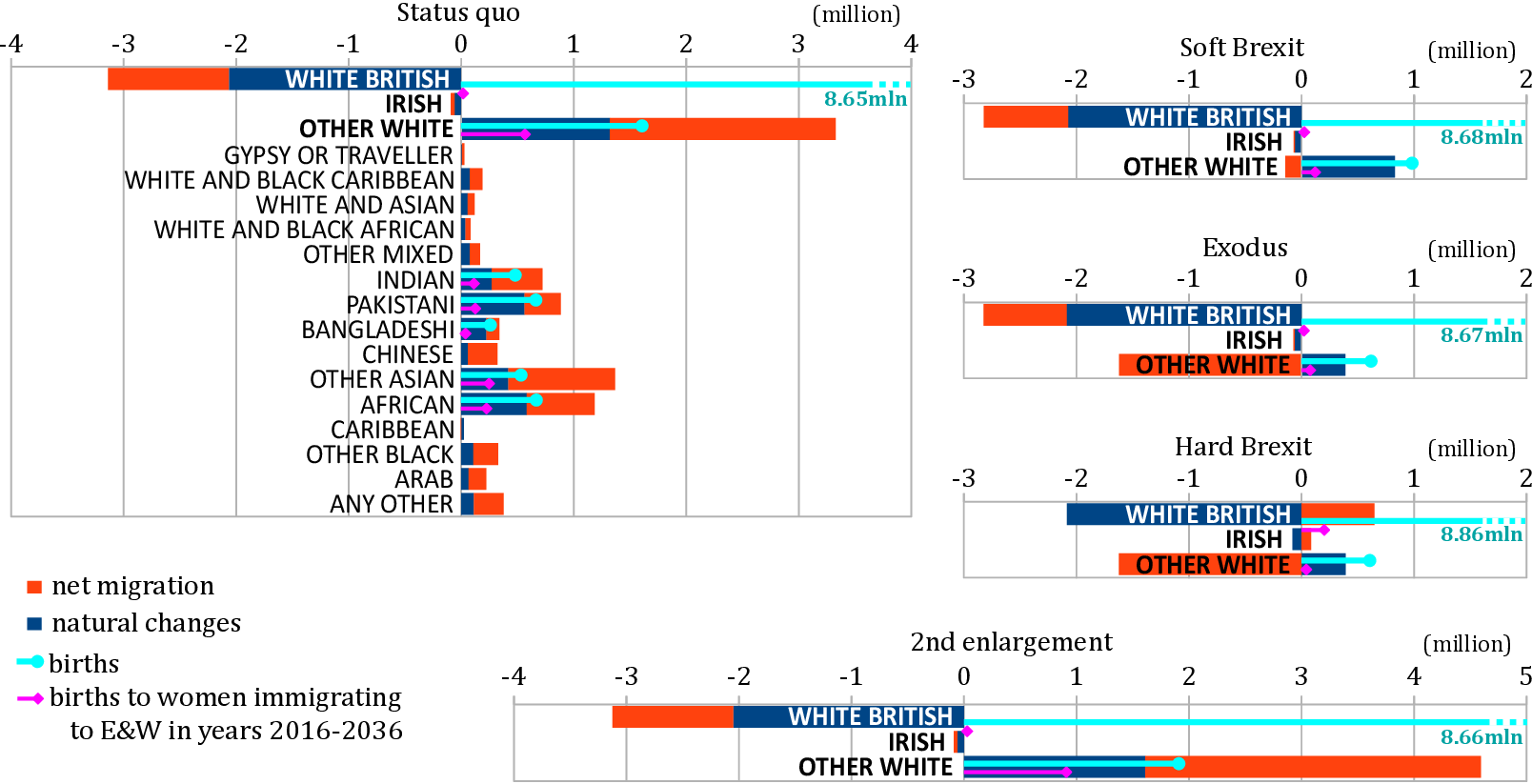}
\caption{Changes of the ethnic group population due to natural growth and net migration between 2016 and 2036 in the E\&W population for different EU membership scenarios. The number of all births (indicated by a number where it exceeds the axis range) and births only to women who immigrate to E\&W in that period are shown for White British and Irish (combined), Other White and several other ethnic groups. The results for ethnic groups unchanged by the scenarios are displayed only in ``Status quo'' chart.}
\label{fig:pop_changes}
\end{figure}

Ethnic composition of the population can change by natural growth and migration. Figure\,\ref{fig:pop_changes} illustrates the contributions of these two factors for the period 2016--2036. In Remain scenarios, the population grows mainly through immigration from the EU and new accession countries after the next enlargement. New immigrant women, who will arrive to E\&W after 2016, contribute 37\% of births to natural growth in this group in the ``Status quo'' scenario and almost 50\% in the ``2nd enlargement'' case. In Brexit scenarios, British expats return, but owing to their relatively small number and predominantly post-reproductive age, they do not counterbalance the negative natural growth of their ethnic group. Their share in the E\&W population shrinks further, even with the diminished presence of the EU immigrants, owing to high fertility rates of several other relatively numerous ethnic groups such as Pakistani or Africans. This above-average fertility rate of non-EU immigrants~\cite{DynamicsDiversity} and the Eastern European ``baby boomers'' (Sec.\,\ref{sec:women-reproductive}) are the main factors driving the population growth.

\section{Conclusions}


We have performed stochastic dynamic microsimulations of E\&W population to forecast its age, sex and ethnic structure under different scenarios of the UK membership in the EU with varying international migration patterns. The differences between Brexit and Remain cases, although significant, are mostly those of degree. In the constantly growing population, the median age and dependency ratios will rise, and the percentage of women of reproductive age will fall regardless of the EU membership status. These changes mostly result from the interplay between longer lifespans (compared with earlier cohorts) and lower average fertility rates, and could be partially forestalled by the influx of immigration from current and prospective EU countries and their children. At the same time, the share of native British in the E\&W population will shrink, even after their repatriation and the exodus of EU immigrants after the Brexit, owing to high fertility rates of other ethnic groups. The only qualitative effects brought about by Brexit will be a temporary population size dip in the ``Exodus'' scenario, a sharp increase of male to female ratio after ``Hard Brexit'' and a significant increase in the median age of EU immigrant group. The above demographic changes will have important fiscal consequences both on the revenue (lower number of people of working age) and expenditure (higher medical care costs and fraction of pensioners offset by lesser demand on school and maternity services) side. Under all considered scenarios, the E\&W demographics will undergo a significant transformation, which can be expected to have a strong impact on the society and politics. Its exact nature will depend on the UK's stance on immigration after Brexit. However, the presented results indicate that alleviating the growing strain placed by the ageing population on the country's social services will require additional interventions. 


\renewcommand{\refname}{REFERENCES}

\end{document}